\documentclass{article}

\usepackage{microtype}
\usepackage{graphicx}
\usepackage{subfigure}
\usepackage{subcaption} 
\usepackage{booktabs} 

\usepackage{hyperref}

\usepackage[accepted]{icml2025}

\usepackage{amsmath}
\usepackage{amssymb}
\usepackage{mathtools}
\usepackage{amsthm}

\usepackage[capitalize,noabbrev]{cleveref}
\usepackage{xcolor}
\usepackage{graphicx}
\usepackage{booktabs, multirow}
\usepackage{enumitem}
\usepackage{makecell}
\usepackage[most]{tcolorbox}
\usepackage{fancyvrb}
\usepackage{fvextra}
\usepackage{float}

\theoremstyle{plain}

\theoremstyle{definition}

\theoremstyle{remark}

\usepackage[textsize=tiny]{todonotes}

\newcommand{\ourmethod}{UI-Evol}

\icmltitlerunning{\ourmethod: Automatic Knowledge Evolving for Computer Use Agents}
\usepackage{float}

\begin{document}

\twocolumn[
\icmltitle{
\ourmethod: Automatic Knowledge Evolving for Computer Use Agents
}



\icmlsetsymbol{equal}{*}

\begin{icmlauthorlist}
\icmlauthor{Ziyun Zhang}{equal,pku}
\icmlauthor{Xinyi Liu}{equal,pku}
\icmlauthor{Xiaoyi Zhang}{msft}
\icmlauthor{Jun Wang}{msft}
\icmlauthor{Gang Chen}{msft}
\icmlauthor{Yan Lu}{msft}
\end{icmlauthorlist}

\icmlaffiliation{pku}{School of Software and Microelectronics, Peking University. Work done during internship in Microsoft Research Asia}
\icmlaffiliation{msft}{Microsoft Research Asia}

\icmlcorrespondingauthor{Xiaoyi Zhang}{xiaoyizhang@microsoft.com}

\icmlkeywords{Agent, ICML}

\vskip 0.3in
]



\printAffiliationsAndNotice{\icmlEqualContribution} 

\begin{abstract}

External knowledge has played a crucial role in the recent development of computer use agents. 
We identify a critical knowledge-execution gap: retrieved knowledge often fails to translate into effective real-world task execution. Our analysis shows even 90\% correct knowledge yields only 41\% success rate.
To bridge this gap, we propose \textbf{\ourmethod{}}, a plug-and-play module for autonomous GUI knowledge evolution. \ourmethod{} consists of two stages: a \textit{Retrace Stage} that extracts faithful objective action sequences from actual agent-environment interactions, and a \textit{Critique Stage} that refines existing knowledge by comparing these sequences against external references.
We conduct comprehensive experiments on the OSWorld benchmark with the state-of-the-art Agent S2. 
Our results demonstrate that \ourmethod{} not only significantly boosts success rate but also addresses a previously overlooked issue of high behavioral standard deviation in computer use agents, leading to superior performance on computer use tasks and substantially improved agent reliability.

\end{abstract}

\begin{figure*}[ht]
\begin{center}
\centerline{\includegraphics[width=\textwidth]{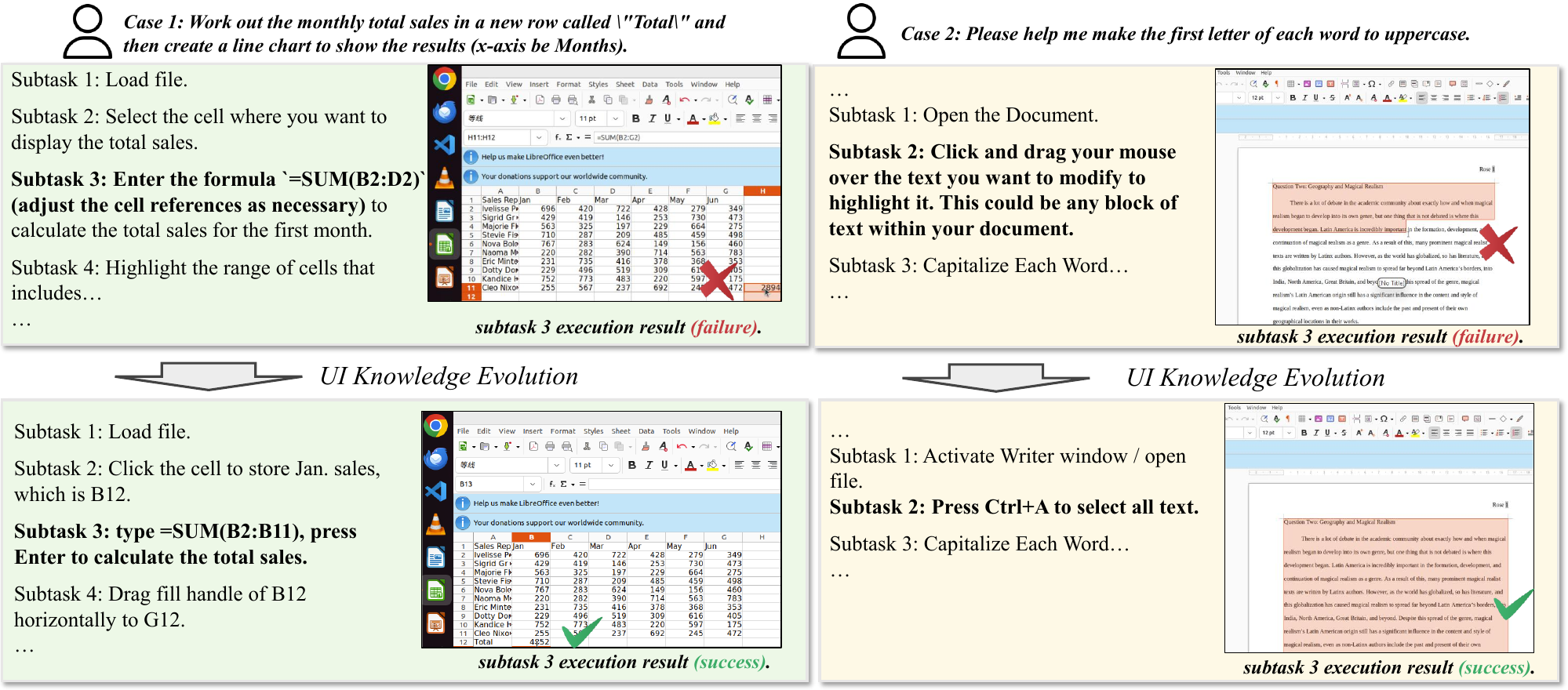}}
\caption{The \textcolor{black}{green box} shows web-retrieved task knowledge, while the \textcolor{black}{yellow box} shows evolved knowledge from our approach. Web knowledge is generally correct but often lacks practical details (left) or suggests with more complex manipulations (right).}
\label{fig:gap_of_knowledge}
\end{center}
\vskip -0.3in
\end{figure*}

\section{Introduction}

Building a computer use agent that can automatically interact with Graphical User Interfaces (GUI) and complete specified tasks with minimal human intervention has always been a major challenge in the field of Autonomous Agents~\cite{huagents,anthropiccua,openaicua}. 
The diversity and heterogeneity inherent in modern GUI interfaces~\cite{commoncrawl} demand that computer use agents exhibit highly robust and generalizable visual perception capabilities. Moreover, the successful long-trace execution of complex, long-horizon tasks further necessitates agents to demonstrate strong reasoning and sequential decision-making capability.
Recent progress in proprietary Large Multimodal Models (LMMs)~\cite{gpt, claude} has substantially advanced the foundational capabilities of computer use agents. This has enabled a surge in research efforts utilizing LMMs as core reasoning engines for automated interface manipulation~\cite{yan2023gpt, zheng2024gpt, agashe2024agent, agashe2025agent}. 

Despite these advancements, it is still challenging
for current LMMs to manipulate software and computer system~\cite{OSWorld,waa,zhouwebarena} based solely on their own knowledge learned in training. 
To mitigate these limitations, some studies~\cite{appagent, appagentv2, oscopilot, skillweaver} have adopted Retrieval-Augmented Generation (RAG)~\cite{lewis2020retrieval} to augment agents with external task-specific knowledge.
By retrieving task-relevant external knowledge, these systems can enhance task planning and execution, alleviating the burden on agents to synthesize execution strategies from scratch for every single scenario~\cite{appagent, appagentv2, oscopilot, skillweaver, agashe2024agent, agashe2025agent}.
For instance, representative Agent S series~\cite{agashe2024agent, agashe2025agent} utilize online web search to obtain existing rich and up-to-date knowledge for given task instruction, including task interpretations and step-by-step plans. Experiments have shown that web-based knowledge can significantly improve agent performance.


Nevertheless, our analysis of Agent S2~\cite{agashe2025agent} based on GPT-4o reveals a persistent gap between the availability of correct knowledge and the knowledge can be effectively consumed by agent for task execution. 
Specifically, in our sampling survey, even when 90\% of the knowledge retrieved via Perplexica~\cite{perplexica} is deemed ``correct'' from the human perspective, the best agent's success rate is only 41\%. More details on the sampling survey are provided in Appendix~\ref{sampling_survey}.
Further investigation into this phenomenon reveals that although the external knowledge may appear theoretically correct and appropriate, it often suffers from certain practical drawbacks, including the omission of necessary intermediate steps (which human might consider as natural), assumptions inconsistent with the initial task conditions, and the suggestion of sub-optimal execution paths demanding overly complex manipulations. 
Figure~\ref{fig:gap_of_knowledge} illustrates two representative cases. In Subtask 3 of Case 1, the external knowledge merely suggests summarizing the column and adjusting the cell reference, which misdirects the agent and leads to task execution failure. Similarly, in Case 2, the provided external web knowledge advises using a click-and-drag action to select text, whereas the task explicitly requires selecting ``all''. Such advice proves challenging for a computer use agent to execute accurately through mouse dragging alone.
These shortcomings underline a critical gap between externally retrieved web knowledge and the actionable knowledge that agents can effectively consume for practical task completion.


To address the aforementioned problem, we propose \ourmethod{}, a plug-and-play module that can be seamlessly integrated into existing computer use agent systems by introducing an autonomous GUI knowledge evolution mechanism aimed at improving knowledge through realistic interactions with practical environments.
Specifically, given a particular task and corresponding initial task knowledge, a computer use agent first performs task operations within the environment, producing an interaction record of actual task execution behaviors.
Subsequently, \ourmethod{} autonomously refines the existing knowledge 
based on actual task execution behaviors, thus mitigating the gap between external knowledge and practical environments.
Specifically, \ourmethod{} consists of two stages:  \textbf{Retrace Stage} and \textbf{Critique Stage}. 
In the Retrace stage, instead of solely relying on the agent’s planned actions for introspective reflection, \ourmethod{} extracts the actual actions executed by the computer use agent based on the screenshots before and after each manipulation. 
These vision-driven observations are then synthesized into a \textit{objective action sequence}, which is a detailed, structural trajectory description that faithfully captures the agent's behavior in the environment, thereby ensuring an accurate and unbiased representation of the agent's concrete interactions with the computer environment.
In the Critique stage, leveraging our earlier observation that retrieved web knowledge is largely reliable, \ourmethod{} uses the externally retrieved knowledge as a reference anchor and further complement it. 
A carefully constructed series of reasoning patterns then assesses and critiques the extracted objective action sequence. 
Specifically, the Critique stage compares the agent-produced action sequences against the reference web knowledge, identifies deviations or anomalies, analyzes underlying causes for these discrepancies, with chain-of-thought reasoning~\cite{wei2022chain} to generate a refined version of task-specific knowledge. This newly evolved knowledge is subsequently stored in a dedicated knowledge base, where it serves as improved reference guidance for subsequent agent executions.



We conduct comprehensive experiments on OSWorld~\cite{OSWorld} to evaluate the effectiveness of \ourmethod{} on state-of-the-art Agent S2. 
The experimental results demonstrate that knowledge evolved by \ourmethod{} is better aligned with an agent in the practical environment.
Notably, during experimentation, we discovered significant instability (high standard deviation) in the baseline computer use agents even when we fix all the hyperparameters we can set, which has been largely overlooked in prior research. 
To systematically examine this instability issue and rigorously evaluate robustness of our approach, we developed a highly efficient parallel evaluation framework.
Utilizing 30 parallel instances, this framework greatly accelerates the agent evaluation, reducing running time from 10 hours to 2.5 hours, achieving approximately 4 times accelerate, thus allowing extensive repetitions of experiments to observe and report the significance of our experiment.
We repeated each experiment variant three times to precisely measure variability. The extensive experimental analyses confirm that \ourmethod{} not only boosts overall success rate but also notably reduces behavioral standard deviation, thus significantly enhancing the robustness and stability of computer use agents.

Our contributions can be summarized as follows:

\begin{itemize}[leftmargin=*,itemsep=2pt, parsep=0pt,topsep=0pt]
  \item We identify the gap between externally acquired knowledge and real task execution, and propose \ourmethod{}, a plug-and-play module that effectively bridges this gap by autonomously evolving GUI task knowledge.
  \item We are the first to systematically identify and analyze the previously overlooked instability issue in contemporary computer use agents, and develop a highly parallelized environment to facilitate efficient investigation.
  \item Comprehensive experiments on the OSWorld benchmark show \ourmethod{} achieves state-of-the-art accuracy and significantly reduces behavioral standard deviation, substantially enhancing agent robustness and stability.
\end{itemize}
\vskip -0.2in

\begin{figure*}[htbp]
\begin{center}
\centerline{\includegraphics[width=0.92\textwidth]{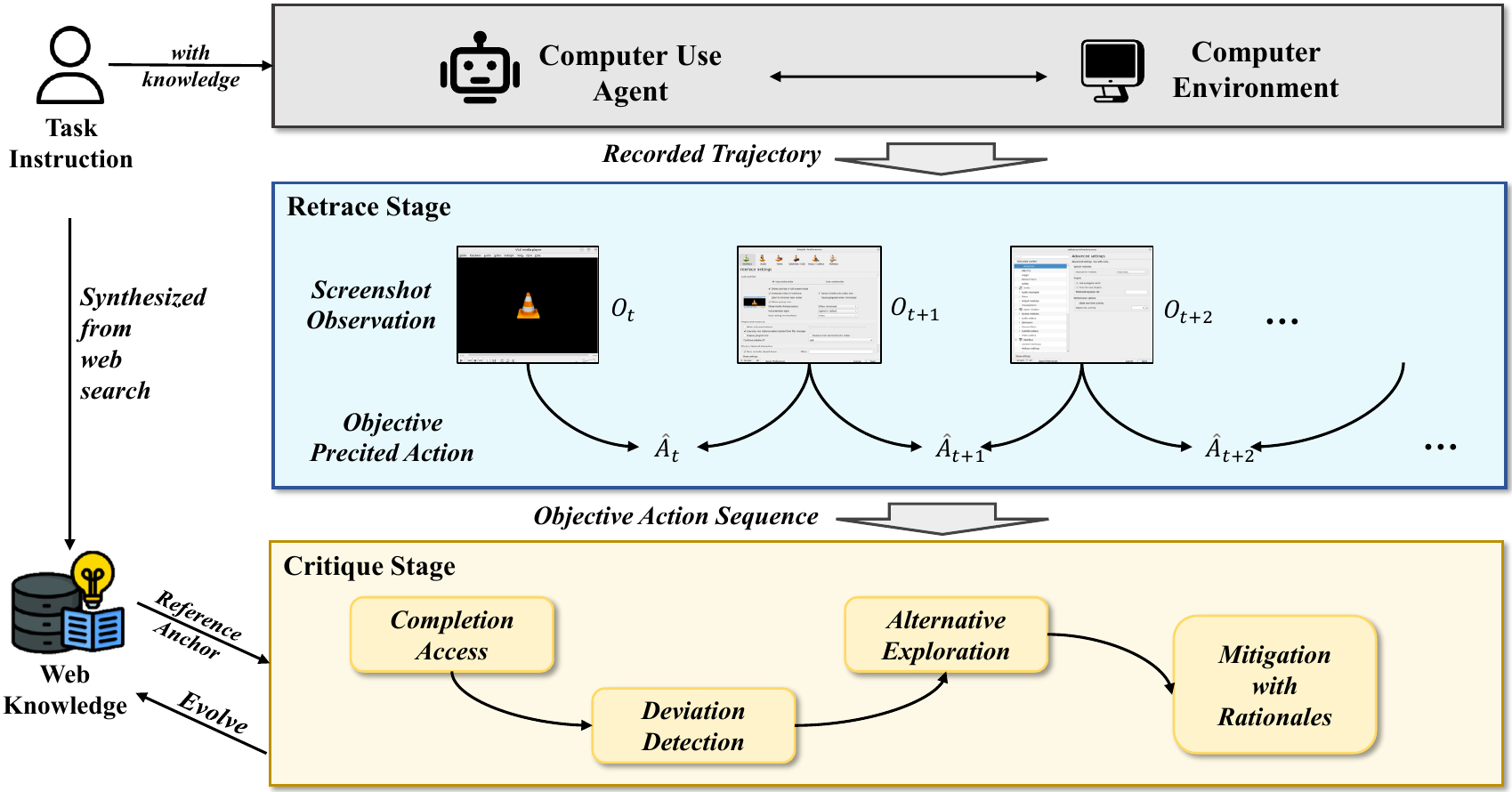}}
\caption{\ourmethod{} consists of two stages: Retrace replays screenshots to recover objective actions; Critique uses web knowledge to detect deviations, explore alternatives, and output rationale-backed fixes that are fed back into the knowledge base.}
\label{fig:overview}
\end{center}
\vskip -0.3in
\end{figure*}

\section{Related Work}

\subsection{Computer Use Agent}
Through the early exploration on automatic computer task execution~\cite{zhang2023responsible,fu2024understanding,seq2act}, recent computer use agents can be broadly categorized into two types: monolithic agents and modular agents. 
A monolithic agent is typically built using a single post-trained, end-to-end model that independently handles the entire computer use task~\cite{zhang2023responsible,cogagent, seeclick, os-atlas, infiguiagent, showui, aguvis, uitars}. Recent open-source pre-trained MLLMs also have native support for computer use via function calls~\cite{qwen25vl}. In addition to supervised fine-tuning, recent works have explored reinforcement learning approaches to train R1-like models~\cite{zhang2023reinforced,xia2025guir1, liu2025infigui, lu2025uir1}. With the increase in both data size and model size, monolithic agents have achieved stronger performance on benchmarks. However, this comes at the expense of high computational demands and scalability limitations. 
Modular agents decompose computer use into multiple modules to reduce the burden on each single model. Some approaches train a separate grounding model and use propriety models to generate actions~\cite{uground, i2e-vlm, yang2024aria}. Some works further divide action generation into multiple stages such as planning and acting, and incorporate external tools for support. For example, some works design multi-level hierarchical planning procedures~\cite{agashe2024agent, wang2024oscar}, while some construct skill libraries to simplify action generation~\cite{skillweaver, appagent, tan2024cradle, oscopilot}. There are also methods that retrieve knowledge from the web to serve as a reference during planning~\cite{agashe2025agent}. However, these additional modules increase the complexity of the overall system and the coupling among them can negatively affect the robustness of the agent.

\subsection{Knowledge refinement and self-evolution}

Early work has demonstrated that LLM-based agents are capable of analyzing failures and summarizing experiences in interactive environments~\cite{wang2023voyager, zhu2023ghost}, thereby enabling self-evolution~\cite{zhou2024symbolic}. A complete self-evolution process typically contains an iterative cycle involving the acquisition, refinement, updating, and evaluation of knowledge~\cite{self-evol-survey}.
Among them, knowledge refinement refers to filtering or correcting knowledge based on environmental feedback, which assists the LLM to adapt to new information and contexts. In environments with factual feedback, some work utilizes such objective signals to guide the refinement of knowledge~\cite{chen2023teaching, zelikman2024star}. More commonly, refinement is driven by an additional critique process, including critiques independently generated by LLM itself~\cite{lu2023self, madaan2023self}, or critiques produced during interactions between LLM and external tools such as code interpreter and Wikipedia~\cite{goucritic, jiang2023selfevolve}. 
Providing appropriate references for the critique process in complex tasks can lead to substantial improvements.
\vskip -0.2in

\section{Preliminary}
\label{sec:howtogenknowledge}
\textbf{Computer use agent with external knowledge.} 
Recent computer use agents typically leverage external knowledge sources to reduce the necessity of generating execution strategies entirely from scratch for each given task. This external knowledge-based approach, which serves as the starting point for our proposed approach, enhances the robustness and success rate of agent execution.
To clearly illustrate how external knowledge is generated and integrated into agent execution, we take Agent S2~\cite{agashe2025agent} as a representative example. 
Specifically, Agent S2 first synthesizes an appropriate query based on the provided task instruction and initial environment state. It then retrieves relevant information via Perplexica~\cite{perplexica}, subsequently summarizing the retrieved information into a structured list of sub-tasks. 
This sub-task list represents the external knowledge format adopted throughout this work.
During task execution, the structured sub-task list is utilized as part of the prompting context provided to the agent, offering a soft prior to assist in efficient and accurate task planning.

\section{Method}
\label{sec:Method}
To mitigate the gap between external knowledge and the actual computer use environment, we propose \ourmethod{}, a plug-and-play retrospective knowledge evolution module for computer use agents. 
Though our method can be applied into any knowledge-based agent, here we take Agent S series as the example to describe our framework.
Following Agent S series~\cite{agashe2024agent, agashe2025agent}, we interpret the web-based knowledge retrieved by agents as a soft prior task plan over the agent's action policy via prompts. 
As depicted in Figure~\ref{fig:overview}, to evolve it, after we obtain the web knowledge as described in Section~\ref{sec:howtogenknowledge}, we  first execute the task instructions to obtain the recorded trajectory.
For the recorded trajectory and web knowledge,
\ourmethod{} consists of two stages: \textbf{Retrace} and \textbf{Critique}. In the Retrace stage, \ourmethod{} reconstructs the agent’s actual trajectory based on screenshots. In the Critique Stage, carefully designed chain-of-thought reasoning patterns are employed to guide the agent in analyzing the causes of deviation and updating the knowledge accordingly.
Without additional human supervision, this process enables the knowledge base to evolve automatically and better support agent execution.
We then introduce the details of the two stages separately.

\subsection{Retrace Stage}

Due to the intrinsic hallucination tendency and limited UI perception capabilities of Large Multimodal Models, existing computer use agents often generate infeasible actions or incorrectly interpret the current computer state. Moreover, the inherent complexity of the computer use environment further exacerbates this issue, leading to executed actions failing to produce the intended effects.
As a result, the action sequences originally output by agents, termed \textit{subjective action sequences}, may not accurately represent actual state changes on the user interface. 
In practice, the subjective action sequence merely represents the intended (rather than actual) behavior of the agent. 
To address the misalignment between the subjective trajectory and the actual state transitions, we introduce the Retrace Stage. In this stage, we propose the Retrace stage to reconstruct an accurate sequence of executed actions, termed \textit{objective action sequences}, based purely on the observed screenshots captured during execution.

Formally, for each step \( t \) in the recorded trajectory, given the observations \( O_t \) and \( O_{t+1} \) before and after the step, the Retrace Stage first enumerates the screenshot in \( O_t \), and then compares the changes in \( O_{t+1} \) relative to \( O_t \). 
A LMM is used to analyze these changes and objectively predict the action $\hat{A}_t$ that occurred during step \( t \). If the LMM thinks nothing happen between \( O_t \) and \( O_{t+1} \),  $\hat{A}_t$ is assigned a null value. The union of all objective actions across steps yields the \textit{objective action sequence}.

We represent the objective action sequence as a sequence of textual descriptions of action, which serves as the input for the next stage. Through the Retrace Stage, the resulting objective trajectory is free from noise introduced by invalid actions, thereby enabling a more reliable comparison with external knowledge sources.

\begin{table*}[ht]
\centering
\small

\begin{minipage}{\textwidth}
\centering
\caption{Performance (\%) of original web knowledge versus \ourmethod{}-refined knowledge across different backbone models on five task groups. SR. denotes the success rate. Std. denotes the standard deviation. Agent S2* is our reproduction result.
}
\label{tab:main_results}
\vskip 0.10in
\resizebox{0.93\textwidth}{!}{

\begin{tabular}{llccccc}  
\toprule  
\textbf{Method} & \textbf{Base Model}  &\textbf{Min. SR.} &\textbf{Max. SR.} &\textbf{Std. SR.} & \textbf{Avg. SR.} & \textbf{Reported SR.} \\  
\midrule  


OpenAI Operator    & OpenAI CUA        &  -   & -    & -    & -    & 19.7      \\
UI-TARS       & UI-TARS-72B-SFT     &  -   & -    & -    & -    & 18.7      \\
UI-TARS       & UI-TARS-72B-DPO     &  -   & -    & -    & -    & 22.7      \\
Aria-UI       & GPT-4o   &  -   & -    & -    & -    & 15.2      \\ 
Aguvis-72B    & GPT-4o   &  -   & -    & -    & -    & 17.0      \\
Agent S2      & GPT-4o   &  -   & -    & -    & -    & 21.1      \\
Agent S2      & Claude-3.7-Sonnet   &  -   & -    & -    & -    & 27.0      \\
\midrule
{Agent S2*}     & GPT-4o      & 18.3          & 20.8          & ±1.00          & 19.5            & 19.5      \\
\ \  \textbf{+ \ourmethod{}}        & GPT-4o      & \textbf{21.0}          & \textbf{22.7}          & ±\textbf{0.71}          & \textbf{22.0}            & \textbf{22.0}\\
{Agent S2*}     & OpenAI-o3   & 24.3          & 27.0          & ±1.09          & 25.6            & 25.6      \\
\ \  \textbf{+ \ourmethod{}}        & OpenAI-o3   & \textbf{28.1} & \textbf{28.6} & \textbf{±0.26} &  \textbf{28.4}  & \textbf{28.4}\\  
\bottomrule  
\end{tabular}

}
\end{minipage}

\vskip 0.15in

\begin{minipage}{0.93\textwidth}
\centering
\caption{Performance (\%) of random vs. completion-based trajectory selection. Rand. Select denotes random selection, and Comp. Select denotes completion-based selection.}
\label{tab:ablation_results}
\vskip 0.10in
\resizebox{\textwidth}{!}{
\begin{tabular}{llcccccccc}  
\toprule  
\textbf{Method} & \textbf{Base Model} &\textbf{OS} & \textbf{Daily} & \textbf{Office} &  
\textbf{Professional} & \textbf{Workflow}  &\textbf{Avg. SR.}  \\  
\midrule  
Ours w/Rand. Select & GPT-4o & 47.22 & 27.83 & 17.98 & 35.61 & 10.43 & 22.0  \\
Ours w/Comp. Select & GPT-4o & 44.45 & 30.39 & 20.54 & 29.96 & 11.43 & 22.7  \\ 
\bottomrule  
\end{tabular}
}
\end{minipage}

\vskip 0.15in

\begin{minipage}{0.93\textwidth}
\centering
\caption{Performance (\%) of transfering evolved knowledge from OpenAI-o3 to GPT-4o.}
\label{tab:migration_results}
\vskip 0.10in
\resizebox{\textwidth}{!}{ 
\begin{tabular}{llcccccccc}  
\toprule  
\textbf{Knowledge Base} & \textbf{Base Model} &\textbf{OS} & \textbf{Daily} & \textbf{Office} &  
\textbf{Professional} & \textbf{Workflow}  &\textbf{Avg. SR.}  \\   
\midrule  
Web Search  & GPT-4o             & 51.39 & 23.98 & 14.27 & 32.43 & 9.11  & 19.5 \\
Evolved from 4o Traj.  & GPT-4o   & 47.22 & 27.83 & 17.98 & 35.61 & 10.43 & 22.0 \\
Evolved from o3 Traj.  & GPT-4o   & 48.61 & 26.12 & 23.09 & 31.33 & 8.78  & 22.4 \\
\bottomrule  
\end{tabular}
}
\end{minipage}

\vskip -0.05in
\end{table*}

\subsection{Critique Stage}


The Critique Stage is designed to refine and systematically enhance the quality of agent knowledge. Our preliminary sampling analysis has demonstrated that the retrieved web-based knowledge generally represents a reliable task-guidance source and thus is leveraged as a reference anchor and further complement it in our framework.
By comparing the reference anchor and the objective action sequence reconstructed during the Retrace Stage, the Critique Stage identifies the gaps between the knowledge and the agent’s actual behavior.
Subsequently, it formulates targeted refinements to directly address and mitigate these discrepancies, thereby systematically evolving the knowledge base toward a closer alignment with the agent’s actual execution policies within real-world task environments. This process results in more effective guidance, greater interpretability, and improved performance for future agent runs.

Specifically, the Critique Stage leverages the chain-of-thought reasoning paradigm, composed of a carefully structured sequence of reasoning steps. Given the \textit{objective action sequence} generated in the Retrace Stage, along with previously acquired web-based knowledge and the task instruction itself, a large language model (LLM) conducts a progressive, multi-stage analysis. This analysis comprises three investigation stages, namely (\textbf{1}) Completion Assessment, (\textbf{2}) Deviation Detection, and (\textbf{3}) Alternative Exploration, followed by a final mitigation stage, namely (\textbf{4}) Mitigation with Rationales. In detail, each analysis step serves a distinct analytical purpose as outlined below:

\begin{itemize}[leftmargin=*,itemsep=2pt, parsep=0pt]
    \item \textbf{Completion Assessment:} This initial step involves assessing whether the agent successfully completed the intended task. Specifically, the LLM compares the outcome depicted in the objective action sequence to the task goal defined in the provided instruction, clearly determining if the task was fully executed or partially completed.
    
    \item \textbf{Deviation Detection:} Subsequently, the LLM conducts a thorough comparative examination between the objective action sequence and the original knowledge-guided action plan. The goal of this step is to explicitly identify deviations, \textit{i.e.}, discrepancies or contradictions between the intended actions outlined by the external knowledge and the actual actions performed by the agent. 
    Crucially, the LLM also infers plausible explanations and root-causes underlying these mismatches, providing deeper insight into systemic failures in action planning or perception that caused the divergence.
    
    \item \textbf{Alternative Exploration:} In this stage, analysis shifts toward understanding the agent’s strategic behavior more comprehensively. The LLM assesses whether the agent, during task execution, attempted valid alternative action strategies beyond or deviating from the original knowledge-based instructions. Identifying such alternative solutions not only yields insights into the robustness and flexibility of the agent's behavior, but also helps to enrich and diversify the evolved knowledge representation.
    
    \item \textbf{Mitigation with Rationales:} Based upon observations, insights, and causal explanations extracted from previous steps, the final mitigation stage synthesizes actionable refinements and corrections. The LLM systematically proposes clear and logically-grounded revisions to the original knowledge base, along with corresponding rationales that explicitly justify changes introduced by this critique process. The resultant output maintains the same representational format as the original knowledge, but is systematically refined and enhanced.
\end{itemize}

Ultimately, upon completing these carefully designed reasoning steps, the LLM generates an updated and refined knowledge representation that explicitly incorporates the identified corrections, clarifications, and supplementary alternative strategies. This renewed knowledge aligns more closely with the agent’s actual behavior, addresses previously discovered inconsistencies, and provides richer, instruction-specific guidance. The evolved knowledge is then recorded and stored in the knowledge base to guide next-round agent executions of analogous tasks.

\section{Experiments}
To demonstrate the effectiveness of our approach, we conducted a series of experiments on OSWorld~\cite{OSWorld}. OSWorld is an interactive environment comprising 369 open-ended computer tasks, where the agent can interact with the environment using screenshots as observations.
We first introduce our parallelization improvements to OSWorld. We then evaluate the performance of \ourmethod{} against the baseline, followed by further ablation studies. All experiments are conducted under maximum 15 steps.

\subsection{Parallel Evaluation Framework}
Because of the inherent stochasticity of large language models, reliably assessing agent performance requires running a large number of repeated trials. However, the original OSWorld benchmark supports only single-machine evaluation and takes about 10 hours to finish even the screenshot-only setting, making large-scale experimentation impractical. To address this limitation, inspired by Windows Agent Arena \cite{waa}, we extend OSWorld by developing a parallel evaluation infrastructure on Microsoft Azure.

Specifically, we employ Azure Machine Learning jobs to parallelize benchmark evaluations across multiple compute instances. Unlike OSWorld, which launches multiple VMWare virtual machines on a single local machine, our implementation automatically provisions one virtual machine per compute instance. Evaluation tasks are evenly distributed across these instances to enable parallel execution. During environment setup, our implementation bypasses repeated Docker image builds that are required in Windows Agent Arena. Instead, it downloads the environment and code directly from the cloud storage. The results are also aggregated in the cloud storage when all experiments finish.

Our implementation enables scalable evaluation of OSWorld, allowing the number of instances to scale up to the total number of benchmark tasks. In our experiments, we used 30 instances, which reduces the full-process runtime of the OSWorld benchmark from approximately 10 hours to 2.5 hours under maximum 15 steps.

\subsection{Main Results}

\paragraph{Settings.}
\label{sec:exp_setting}
For our experiments, we adopt GPT-4o and OpenAI-o3 as the base models for Retrace Stage and Critique Stage separately. We first run computer use agent with web knowledge to obtain the recorded trajectory and the knowledge to be evolved later.
Then we run our \ourmethod{} to evolve the knowledge.
Finally we equip the same computer agent with the updated knowledge to evaluate the improvement brought by the knowledge evolution.
We select Agent S2~\cite{agashe2025agent} as our baseline, which is the leading computer use agent on OSWorld. Agent S2 uses Perplexica for web search and references the retrieved external knowledge to decompose user instructions into multiple subtasks, which are then executed by a coordinated set of agents. Since Agent S2 retrieves web knowledge at the beginning of each run, we capture and freeze a snapshot of the entire knowledge base before the first experiment while set all hyperparameters as constant value including temperature as 0. This eliminates variability caused by dynamic web.

\paragraph{Stability.} During experiments, we observe that even with static precaptured web knowledge and fixed hyperparameters, repeated evaluations may still yield different results. We argue that producing stable and reproducible outcomes is a critical capability for agents, especially for real-world deployment. Therefore, we include stability as one of our evaluation criteria. For each experimental variant, we repeat the same run three times and compute the mean and standard deviation of the success rate. The mean reflects the agent's overall performance, while the standard deviation indicates the agent’s stability.



\begin{figure*}[htbp]
\vskip -0.2in
\begin{center}
\centerline{\includegraphics[width=\textwidth]{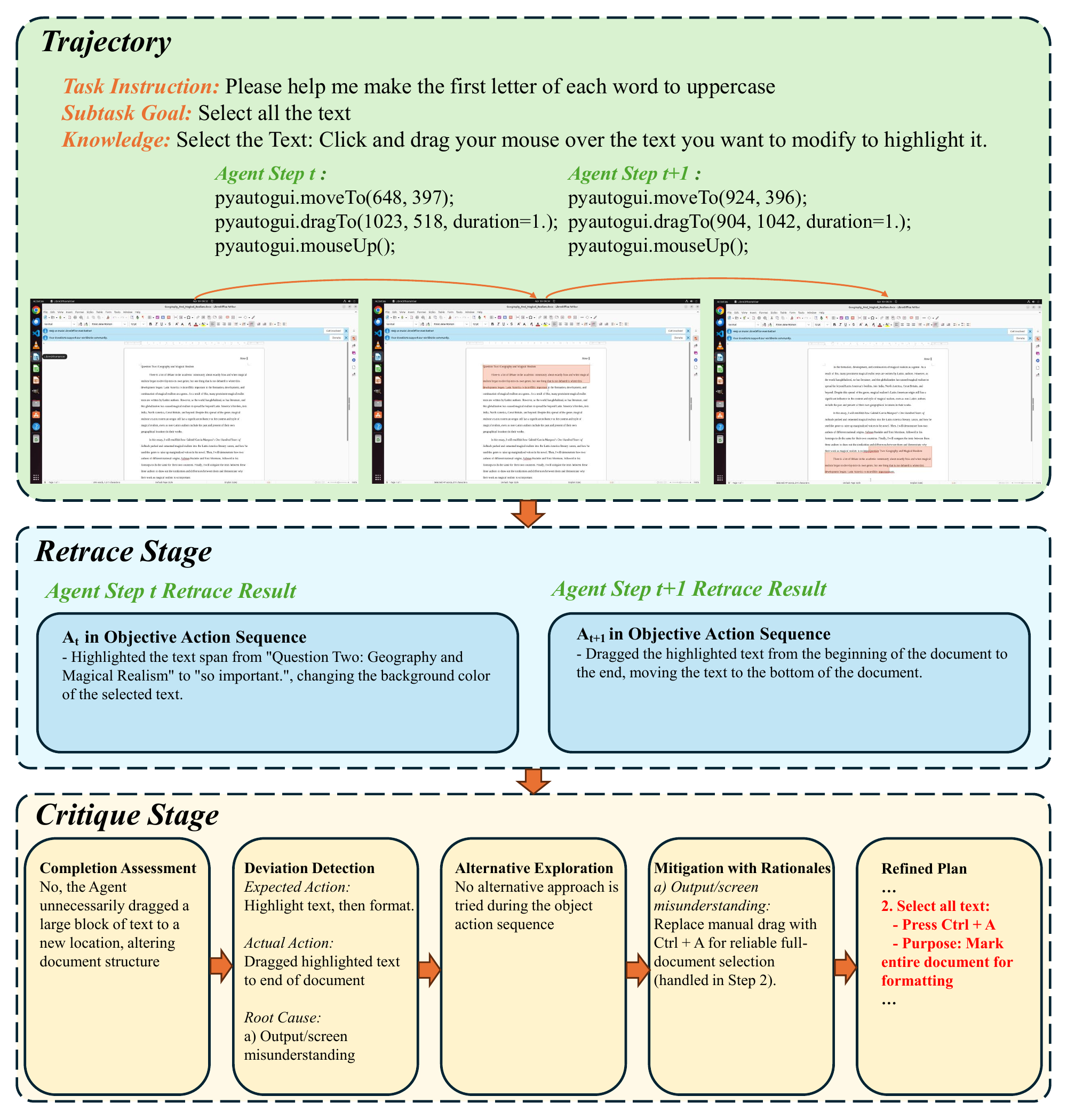}}
\caption{Case study on the ``capitalize every word'' task from the OSWorld benchmark: Our \ourmethod{} first retraces the objective action sequence from the screenshots and identifies that the action taken at step $t$ was selecting only a part of the paragraph. In the Critique Stage, it detects that this action deviates from the objective, as the entire document should have been selected rather than a partial selection. Finally, our framework corrects this deviation by proposing a simpler keyboard shortcut, ``Ctrl + A'', instead of dragging with the mouse.}
\label{fig:case_study}
\end{center}
\vskip -0.2in
\end{figure*}

\paragraph{Results}
Using Agent S2 as the baseline, we compare the performance of using only the original web search knowledge with that of using knowledge refined by~\ourmethod{} across different base models. As reported in Table~\ref{tab:main_results}, \ourmethod{} consistently improves the average success rate and reduces standard deviation on both models. This demonstrates that~\ourmethod{} can effectively enhance both agent performance and stability.
Notably, when integrating~\ourmethod{}, the standard deviation of OpenAI-o3 drops to as low as 0.26, approximately 4.19 times the reduction observed with GPT-4o. This suggests that models with stronger reasoning capabilities are better at understanding and leveraging external knowledge, and therefore benefit more from~\ourmethod{}.

\subsection{Ablation Study}
As mentioned in Section~\ref{sec:exp_setting}, we repeat each run for three times during evaluation. Due to the stochastic nature of LLMs, the trajectories generated in each run may differ. To study how the selection strategy affects the final results, we compare two approaches: \textbf{(1)Random selection}, where a trajectory is randomly chosen from the repeated runs; \textbf{(2) Completion-based selection}, where all trajectories are first transformed to textual format through the Retrace Stage, and then an LLM is prompted to select the trajectory from completeness. We detail the prompt in our appendix.


To measure the quality of selection, we introduce the Selection Success Rate (SSR) metric, defined as \(\mathrm{SSR} = \frac{N_{\text{succ}}}{N_{\text{solv}}}\), where \(N_{\text{succ}}\) is the number of cases where the selected trajectory (\(\tau_i\)) is successful, and \(N_{\text{solv}}\) is the number of cases in which at least one out of three repeated runs succeeds. Here, \(\tau_i\) denotes a trajectory generated in an experiment. A higher SSR indicates that the completeness of trajectories.


Under this metric, random selection  achieves an SSR of 70\%, while completion-based selection reaches 85\%. This demonstrates that the LLM can identify trajectories that are more likely to be correct. We use trajectories selected by these two approaches as inputs to \ourmethod{}, and compare their performance using GPT-4o as the base model. As shown in Table \ref{tab:ablation_results}, \ourmethod{} with the trajectories chosen based on completeness performs only marginally better. 
These results suggest that \ourmethod{} is robust to the quality of trajectory selection and can effectively leverage knowledge from both better and worse inputs.

\subsection{Knowledge Transfer}
To evaluate whether knowledge evolved from one model can be effectively reused by another, we use the OSWorld trajectories generated by OpenAI-o3 to derive refined knowledge through \ourmethod{}.
This knowledge is then provided as input to Agent S2 with GPT-4o, as shown in Table~\ref{tab:migration_results}. 
Compared to the original GPT-4o baseline which leverages web knowledge, the agent using refined knowledge derived from OpenAI-o3 trajectories performs similarly to that of using GPT-4o’s own trajectories. These findings demonstrate that the knowledge refined through \ourmethod{} can indeed be transferred across different models, capturing task regularities.

\subsection{Case Study}
\paragraph{Case introduction.}
In the task of Figure~\ref{fig:case_study}, the agent is required to capitalize the first letter of each word in a LibreOffice Writer file. The first step is to select all the words in the document. From web knowledge, the agent is instructed to select entire text by clicking and dragging with the mouse. However, despite strictly following the guidance of the knowledge, the agent only selected part of the text in the first attempt and dragged the selected text to the end of the passage in the second, ultimately failing the task.
\paragraph{Evolution progress.}
After the trajectory is fed into \ourmethod{}, the system first enters the Retrace Stage, where each step is distilled into two textual elements: Action and Result. In the first Retrace step, \ourmethod{} analyzes the pre-execution screenshot and summarizes the initial state, correctly noting that the agent selected only the section from “Question Two: Geography and Magical Realism” to “so important.” In the second Retrace step, \ourmethod{} precisely describes both the outcome of that action and the agent’s error—dragging the selected text from the beginning to the end of the passage.

With this objective record in hand, the Critique Stage evaluates the trajectory. During Completion Assessment it observes that the task failed because the drag operation altered the document’s structure. In Deviation Detection it traces the failure to an error in the selection step and labels it an Output/Screen Misunderstanding. The Alternative Exploration stage finds no evidence that the agent attempted other methods. In the Mitigation with Rationales stage, it recommends replacing manual dragging with the more reliable Ctrl + A shortcut, noting that this change will appear in step 2 of the final plan.

Finally, drawing on all prior analyses, \ourmethod{} revises step 2 accordingly, thereby completing the learning pipeline and incorporating the new knowledge into its repertoire.

\subsection{Computational Cost and Scalability}
As detailed in Section~\ref{sec:Method}, \ourmethod{} consists of an initial raw-trace collection phase followed by two sequential stages: Retrace and Critique. During the raw-trace collection phase, 30 OSWorld instances process jobs in parallel, requiring approximately two hours to generate the raw traces. The subsequent Retrace and Critique stages are executed on 12 parallel threads and together process all 369 OSWorld tasks in roughly one hour.

The primary monetary cost stems from OpenAI API usage. In the Retrace stage, each 15-step trace consumes around 85000 input tokens and 400 output tokens and is processed by the GPT-4o model. In the Critique stage, each trace requires around 800 input tokens and 150 output tokens and is handled by the OpenAI-o3 model. On average, each task incurs a cost of \$0.22, resulting in a total expenditure of approximately \$81.18 for the entire benchmark.

For scalability, \ourmethod{}'s task-level pipeline is highly parallelizable: adding compute instances yields near-linear throughput gains while maintaining per-task latency, ensuring efficient development iteration at scale.

\section{Conclusion}

In this work, we tackle the knowledge-execution gap in computer use agents. We propose \ourmethod{}, an autonomous, plug-and-play module that evolves GUI interaction knowledge by evolving external knowledge based on actual agent behaviors. Our experiments on OSWorld show \ourmethod{} significantly improves task accuracy and, crucially, reduces agent behavioral variance, enhancing stability. This work offers a possible solution towards more reliable and advanced autonomous computer use.



\section*{Impact Statement}
Computer use agents hold immense potential to significantly boost human productivity on digital devices, yet their misuse poses considerable risks to society. For instance, these agents could be exploited for malicious purposes, such as the automated creation of numerous spam accounts, leading to a deluge of unwanted content and potential security breaches.

\nocite{langley00}

\bibliography{example_paper}
\bibliographystyle{icml2025}

\newpage
\appendix
\onecolumn

\section{Domain Specific Results on OSWorld}

\begin{table}[ht]  
  \caption{Performance (\%) of different settings across individual applications and overall score on GPT-4o.}  
  \label{tab:appendix_results_4o}  
  \vskip 0.10in  
  \centering  
  \small    
  \resizebox{\textwidth}{!}{%
  \begin{tabular}{lcccccccccccc}  
    \toprule  
    \textbf{Method} & \textbf{Chrome} & \textbf{Gimp} & \textbf{Calc} & \textbf{Impress} &  
    \textbf{Writer} & \textbf{Multiapps} & \textbf{OS} & \textbf{TB} &  
    \textbf{VLC} & \textbf{VSCode} & \textbf{Avg. SR.}\\  
    \midrule  
Agent S2*                & 22.31 & 16.66 & 7.09 & 17.10 & 23.18 & 9.11 & 51.39 & 35.56 & 18.30 & 46.38 & 19.49 \\
 \textbf{+ UI-Evol w/Rand. Select}     & 25.12 & 21.80 & 9.22 & 20.64 & 30.43 & 10.43 & 47.22 & 44.44 & 20.27 & 47.83 & 22.02 \\
 \textbf{+ UI-Evol w/Comp. Select}     & 27.38 & 17.95 & 11.35 & 18.52 & 43.47 & 11.43 & 44.45 & 44.44 & 26.14 & 40.58 & 22.74 \\
 \textbf{+ UI-Evol w/knowledge evolved from o3}  & 21.02 & 19.23 & 16.31 & 26.29 & 30.43 & 8.78 & 48.61 & 44.45 & 23.76 & 42.03 & 22.38 \\
    \bottomrule  
  \end{tabular}  
  }  
  \vskip -0.05in  
\end{table}  

\begin{table}[ht]  
  \caption{Performance (\%) of different settings across individual applications and overall score on OpenAI-o3.}  
  \label{tab:appendix_results_o3}  
  \vskip 0.10in  
  \centering  
  \small    
  \resizebox{\textwidth}{!}{%
  \begin{tabular}{lcccccccccccc}  
    \toprule  
    \textbf{Method} & \textbf{Chrome} & \textbf{Gimp} & \textbf{Calc} & \textbf{Impress} &  
    \textbf{Writer} & \textbf{Multiapps} & \textbf{OS} & \textbf{TB} &  
    \textbf{VLC} & \textbf{VSCode} & \textbf{Avg. SR.}\\  
    \midrule  
Agent S2*   & 28.21 & 20.51 & 19.15 & 24.19 & 40.57 & 9.11 & 55.55 & 51.11 & 31.82 & 47.83 & 25.64 \\ 
 \textbf{+ UI-Evol}    & 29.71 & 32.05 & 24.82 & 25.59 & 44.92 & 12.42 & 55.56 & 42.22 & 32.07 & 46.38 & 28.28  \\
    \bottomrule  
  \end{tabular}  
  }  
  \vskip -0.05in  
\end{table}

\section{Sampling Survey On Web Knowledge}
\label{sampling_survey}
To evaluate the reliability of knowledge retrieved from web via Perplexica, we conduct a sampling survey. Specifically, we sampled 50 test cases proportionally across various domains from OSWorld, and manually assess the correctness of the knowledge retrieved in each case. The knowledge is considered ``correct'' if an annotator can successfully complete the given task using only that information without relying on any prior domain knowledge. Under this criterion, 45 of the 50 sampled cases are deemed to contain correct knowledge from the human perspective. 

\section{Prompts}

\begin{tcolorbox}[colback=white, 
    colframe=gray!50!black, 
    coltitle=black, 
    title=\textbf{Prompts for Retrace Stage}, 
    fonttitle=\bfseries\large, 
    colbacktitle=white!80!gray, coltitle=black,
    left = 1mm,right = 1mm,top = 1mm,bottom = 1mm,
    enhanced, 
    breakable=true,
    after skip=0pt,
    ]
\footnotesize
{
\begin{Verbatim}[breaklines=true]
You are a senior QA assistant.
You receive:
• BEFORE screenshot <image0>    
• AFTER  screenshot <image1>    
• A snippet of Python automation code.    

Your task:    

PART A - BEFORE DESCRIPTION    
Describe concisely and objectively what is visible in the BEFORE screenshot only.    
• <= 80 words, declarative sentences.    
• No speculation, no mention of AFTER, no hidden reasoning.    

PART B - UI OPERATION LIST    
List, in chronological order, every visible UI step (mouse-click, key-stroke, drag, menu selection…) that converted the BEFORE state into the AFTER state.    

OUTPUT FORMAT (STRICT)    
[A] BEFORE    
<one-to-three short sentences that satisfy PART A>    

[B] OPERATIONS    
- <action>, <visible consequence>    
- …    

RULES FOR PART B (inherited)    
1. Bullet list; every line begins with "-".    
2. Each bullet MUST pair the action with its visible consequence, e.g.    
- Clicked the "Replace All" button in VS Code's Search sidebar, replacing all 12 occurrences of "text" with "test" in the open file    
3. Do not add headings, explanations or blank lines beyond the specified format.    
4. If the ONLY difference is the system clock, Part B must contain exactly one bullet:    
- No operations performed.    
5. If the screenshots cannot be compared, Part B must contain exactly one bullet:    
- Unable to determine operations.    

Think step-by-step internally but reveal ONLY the two required sections.    

FEW-SHOT EXAMPLES    

<BEGIN_EXAMPLE>    
# Normal change with visible result    
BEFORE: VS Code shows 3 occurrences of "foo"    
AFTER : All occurrences now read "bar"    
CODE  : editor.replace_all("foo", "bar")    
OUTPUT:    
[A] BEFORE    
VS Code editor window is open; the Find/Replace panel indicates 3 matches for the word "foo".    

[B] OPERATIONS    
- Pressed Ctrl+H in the VS Code editor, opening the Find/Replace panel    
- Typed "foo" into the Find box, highlighting 3 matches in the file    
- Typed "bar" into the Replace box    
- Clicked the "Replace All" button in the Find/Replace panel, replacing all 3 occurrences of "foo" with "bar" in the document    
<END_EXAMPLE>    

<BEGIN_EXAMPLE>    
# Only the clock changed    
BEFORE: Desktop 10:01    
AFTER : Desktop 10:02    
OUTPUT:    
[A] BEFORE    
Desktop environment showing wallpaper and system clock reading 10:01.    

[B] OPERATIONS    
- No operations performed.    
<END_EXAMPLE>    

<BEGIN_EXAMPLE>    
# Incomparable    
BEFORE: Corrupted screenshot    
AFTER : Corrupted screenshot    
OUTPUT:    
[A] BEFORE    
Screenshot is corrupted; no discernible UI elements are visible.    

[B] OPERATIONS    
- Unable to determine operations.    
<END_EXAMPLE>

The FIRST image (<image0>) shows the screen BEFORE the Agent acted.  
The SECOND image (<image1>) shows the screen AFTER the Agent acted.  

The Agent executed the following Python code:  

```python  
{code}    
List the UI operations (action + visible result).

\end{Verbatim}
}
\end{tcolorbox}

\begin{tcolorbox}[colback=white, 
    colframe=gray!50!black, 
    coltitle=black, 
    title=\textbf{Prompts for Critique Stage}, 
    fonttitle=\bfseries\large, 
    colbacktitle=white!80!gray, coltitle=black,
    left = 1mm,right = 1mm,top = 1mm,bottom = 1mm,
    enhanced, 
    breakable=true,
    after skip=0pt,
    ]
\footnotesize
{
\begin{Verbatim}[breaklines=true]
INPUT  
  Task Instruction: …  
  Action List: …  
  Original Plan: …  

REQUIREMENTS  
    • Follow the FIVE SECTION HEADERS below exactly.  
    • SECTION E output style:  
          1. **<Subtask>**:  
             - <Concrete UI / CLI action(s) only>  
             - Purpose: <<= 10-word reason>  
    • If a field is not applicable, write “None” or “No deviation”.  
    • If SECTION C judges an Alternative better, the final NEW PLAN must adopt it (or its key advantages).  
    • Every Root Cause from SECTION B must have a mitigation explained in SECTION D and be implicitly addressed (not as a standalone step) in SECTION E.  
    • Exclude passive “Confirm / Verify / Check / Make sure …” kinds of steps.  
    • Visual inspections are assumed; do not list them.
    • If the Action List shows a dialog / branch / extra option that the Original Plan did not anticipate:
    - Treat it as a Deviation (Root Cause usually f) Invalid assumption).
    - If the Agent picked the wrong option, SECTION D must state the correct option and SECTION E must insert that corrected step.
    - If the Agent picked the right option, still add that step to SECTION E (it is an “added step”).
    - Any action shown to be unnecessary in the trajectory must be omitted from SECTION E (this is a “removed step”).

SECTION A. Task Completion  
  Did the Agent achieve the task goal? (Yes / No)  
  Reason.
  Did the Agent execute more than the instruction required? (Yes / No)
  Reason.

SECTION B. Deviation Analysis  
  For every mismatch between an Original-Plan assumption and the actual screen/CLI output in Action List, record a Deviation row. Fill in ALL items, even if “No deviation”.  
  • Deviation Step: <# or “None”>  
  • Expected Action : …  
  • Actual Action    : …  
  • Root Cause (letters, commas allowed):  
        a) Output/screen misunderstanding  
        b) Knowledge gap  
        c) Command / code / syntax error  
        d) Environment or permission issue  
        e) Other  
        f) Invalid assumption  
        g) External transient failure
        h) Step order issue
        i) Missing precondition

SECTION C. Alternative Approaches  
  Did the Agent attempt any approach beyond the Original Plan? (Yes / No)  
  If Yes:  
      • Describe each approach briefly.  
      • Which is better (Original / Alternative)? Why?  
  If No: “No alternative approach tried.”  

SECTION D. Mitigation & Rationale  
  For every Root Cause from SECTION B, describe the preventive or corrective idea and mention which forthcoming step embodies it.  
  Example:  
      c) Syntax error → Add “lint before run” check (handled in Step 2).  
      d) Permission → Verify sudo rights before executing installer (Step 5).
      f) Invalid assumption → Choose “Typical” in installer dialog (Step 2).  

SECTION E. REFINED PLAN:  
  REFINED PLAN:  
      1. **<Subtask>**:  
         - <Concrete action(s)>  
         - Purpose: <Why this step?>  
      2. **<Subtask>**:  
         - <Concrete action(s)>  
         - Purpose: …  
      …  
      up to 15 steps total.  
  • No shell prompts (#, $).  
  • Safeguards are implicit per SECTION D; do not list them as separate lines.
  • Newly added corrective steps must appear in the proper sequence among Steps 1-15; actions deemed unnecessary must not appear here.

\end{Verbatim}
}
\end{tcolorbox}

\begin{tcolorbox}[colback=white, 
    colframe=gray!50!black, 
    coltitle=black, 
    title=\textbf{Prompts for Completion-based selection}, 
    fonttitle=\bfseries\large, 
    colbacktitle=white!80!gray, coltitle=black,
    left = 1mm,right = 1mm,top = 1mm,bottom = 1mm,
    enhanced, 
    breakable=true,
    after skip=0pt,
    ]
\footnotesize
{
\begin{Verbatim}[breaklines=true]
INPUT:
1. Task Instruction   : The instruction for the task.
2. Action_List1: The list1 of actions performed by a linux user.
3. Golden_Plan1: The plan1 that the user is trying to achieve to solve a task.
4. Action_List2: The list2 of actions performed by a linux user.
5. Golden_Plan2: The plan2 that the user is trying to achieve to solve a task.
6. Action_List3: The list3 of actions performed by a linux user.
7. Golden_Plan3: The plan3 that the user is trying to achieve to solve a task.

REQUIREMENTS:
You need to find out the best action_list and golden_plan pair that is most likely completed or closest to completion.
Your output should include the following information:
- <Analysis and Score>: Analysis and Give your score(0 to 10) for each pair, 10 is the best, 0 is the worst.
- <Best Pair>: A number in [1, 2, 3] indicating the best action_list and golden_plan pair.

\end{Verbatim}
}
\end{tcolorbox}

\end{document}